\documentclass{article}

\usepackage{spconf_modified,amsmath,epsfig}
\usepackage{hyperref}

\usepackage{setspace}
\usepackage{latexsym}
\usepackage{amsfonts}
\usepackage{curves}
\usepackage{epic}
\usepackage{eepic}
\usepackage{epsfig}
\usepackage{url}
\usepackage{psfrag}
\usepackage{subfigure}
\usepackage{times}
\usepackage{color}
\usepackage{helvet}
\usepackage{caption}
\usepackage{courier}
\usepackage{amsmath, amsthm, amssymb}
\usepackage{graphicx}

\usepackage{cite}
\usepackage{subfigure}

\usepackage{xspace}
\newcommand{\dynamicEnergy}{radiated energy\xspace} 
\newcommand{\staticEnergy}{operating energy\xspace} 

\newtheorem{proposition}{Proposition}

\newtheorem{example}{Example}

\newtheorem{definition}{Definition}

\newtheorem{remark}{\bf Remark}

\newtheorem{fact}{Fact}

\newcommand{\setn}{\mathcal{N}}
\newcommand{\setm}{\mathcal{M}}

\newcommand{\signal}[1]{{\boldsymbol{#1}}}

\newcommand{\Natural}{{\mathbb N}}

\newcommand{\real}{{\mathbb R}}

\newcommand{\refeq}[1]{(\ref{#1})}

\usepackage{enumerate}

\newcommand{\cosl}[1]{}
\newcommand{\cosr}[1]{}
\newcommand{\insl}[1]{#1}

\usepackage{cancel}
\usepackage{soul}
\newcommand{\resl}[1]{}
\usepackage{fancybox}

\let\OLDthebibliography\thebibliography
\renewcommand\thebibliography[1]{
  \OLDthebibliography{#1}
  \setlength{\parskip}{0pt}
  \setlength{\itemsep}{0pt plus 0.3ex}
}

\title{Toward Energy-Efficient 5G Wireless Communications Technologies}%
\name{Renato L. G. Cavalcante$\dagger$, S\l awomir Sta\'nczak$\dagger\star$, Martin Schubert$\ddagger$, Andreas Eisenbl\"atter$\ast$, Ulrich T\"urke$\ast$ \thanks{Copyright (c) 2014 IEEE. Personal use of this material is permitted. However, permission to use this material for any other purposes must be obtained by sending a request to pubs-permissions@ieee.org. The work was partially supported by the European Commission within the FP7 Projects GreenNets (grant agreement n.~286822) and by the German Federal Ministry of Economics and Technology (BMWi) within the project ComGreen (grant number 01ME11010). Part of this work has also been performed in the framework of the FP7 project ICT-317669 METIS, which is partly funded by the European Union. The authors would like to acknowledge the contributions of their colleagues in METIS, although the views expressed are those of the authors and do not necessarily represent the project. }}%
\address{$\dagger$ Fraunhofer Heinrich Hertz Institute, Berlin, Germany. $\quad\star$ Technical University of Berlin, Germany \\ $\ddagger$ Huawei Technologies - European Research Center, Munich, Germany $\ast$ atesio GmbH, Berlin, Germany}
\begin{document}
\maketitle
%
\begin{abstract}
  The densification and expansion of wireless networks
  pose new challenges on energy efficiency. With a drastic increase of
  infrastructure nodes (e.g. ultra-dense deployment of small cells),
  the total energy consumption may easily exceed an acceptable level.
  While most studies focus on the energy radiated by the antennas, the
  bigger part of the total energy budget is actually consumed by the
  hardware (e.g., coolers and circuit energy consumption). The ability
  to shutdown infrastructure nodes (or parts of it) or to adapt the
  transmission strategy according to the traffic will therefore become
  an important design aspect of energy-efficient wireless architectures.
  Network infrastructure should be regarded as a resource that can be
  occupied or released on demand, and the modeling and optimization of such systems are highly nontrivial problems. In particular, elements of the network infrastructure should be released by taking into account traffic forecasts to avoid losing the required coverage and capacity. However, even if traffic profiles were perfectly known, the determination of the elements to be released is complicated by the potential interference coupling between active elements and the sheer size of the optimization problems in dense networks.

\end{abstract}

\section*{\small INTRODUCTION}

Due to the compelling need for broadband mobile access to the
Internet, over the past decade there has been a dramatic growth in
demand for wireless access worldwide. This growth is expected to
continue in the years to come, driven by an increasing interest in
various wireless services and novel types of machine-to-machine (M2M)
and device-to-device (D2D) communications. The vision is to create the
so-called ``Internet-of-Things'' by integrating billions of sensors
and actuators into physical objects and connecting them to the network
via wireless connections. Requiring no human involvement, such
communications may exceed any existing limits on information
dissemination, leading to a data explosion of unprecedented magnitude.

The above vision can only be brought to reality if we introduce major
changes to the way current cellular networks are designed and
operated. The need for such changes can be partially justified by the
results in the landmark study by Gupta and Kumar
\cite{Gupta2000}. {This study
  strongly suggests} that {traditional} large-scale networks
{(i.e., networks of spatially and temporarily independent
  sources, arbitrarily located source and destination nodes, and
  arbitrary traffic demands)} inevitably face the problem of
{asymptotically} vanishing {per-user} throughputs whenever the
following restrictions hold: i) {nodes} are
stationary, ii) interference is treated
as noise {\cite{slawomir09}}, and iii) the network operates
without any underlying infrastructure such as the presence of base
stations (i.e., the information needs to be carried from node to node
in a multi-hop fashion). As an immediate consequence, to overcome this
fundamental limit, we must drop at least one of the restrictions when designing wireless networks in order to provide additional dimensions for
network optimization. \par

The first restriction of the study in \cite{Gupta2000} has been lifted
in \cite{grossglauser:2002}, where the authors analyze an
infrastructure-less network with mobile {nodes} and
interference treated as noise. It has been shown that mobility can
stabilize the throughput at the cost of delay. Further studies reveal
and characterize a fundamental trade-off between throughput and delay
\insl{(e.g., \cite{neely:2005})}. {From these
  studies,} we conclude that mobility {may}
be an important ingredient {in enhancing} the
performance of future wireless networks, but, {due to strict
  delay constraints of many wireless applications}, it \resl{is
  not} {cannot be} the ultimate solution {to the problem.}

By dropping the second restriction, we can make nodes exploit, shape,
or reject interference through advanced multi-user transmission and
reception techniques. Indeed, the studies in \cite{Xie04},
{\cite{Ozgur07} have shown that the
throughput in large-scale infrastructure-less networks can be
stabilized by resorting to cooperation and
{other} multi-user communications strategies that
are derived from information-theoretic results on broadcast,
multiple-access, and relay channels. An additional
performance-enhancing approach falling into the class of
interference-shaping techniques is interference
alignment~\cite{CadambeJafar08}. However, these
interference-mitigating methods alone cannot deliver the promised
gains in practice because of, for example, the lack of channel state
information and the lack of perfect synchronization in real
systems. Therefore, a fixed network infrastructure, which is of vital
importance to current networks, is also envisioned to play a crucial
role in future systems.

In particular, networks with densely deployed infrastructure nodes are
one of the main pillars in the current 5G discussion to enhance the
throughput of cellular networks at relatively low operational costs
\cite{HoyKobDeb11}. The vision is to have small and low cost
base stations to form small cells and to provide Internet access by
using short-distance links
\cite{METIS}. This vision is partially motivated by the analysis in
\cite{Gupta2000}{; in fact, the study in \cite{liu:2003}
  shows} that the per-user throughput can
{be improved significantly} even when interference
is treated as noise, provided that the density of
infrastructure nodes grows sufficiently fast with the number of
users.

One of the main challenges that may limit the acceptable density of
future networks is the high capital and operational costs. In
particular, a large part of operational costs is directly related to
the energy consumed for transmission and for operation of the network
infrastructure~\cite{correia10}. We argue in this article that future
5G wireless communications technologies need to be energy-efficient to
reduce the total cost per transmitted bit, thereby providing
cost-effective, affordable wireless bandwidth.

In the text that follows, we give an overview of fundamental aspects
of energy saving mechanisms. Using information-theoretic arguments, we
show that, without any additional energy saving measures, current
approaches aiming at reducing the (radiated) energy-per-bit may, in
fact, increase the {\it total} energy consumption because of, for
example, the energy required to power hardware and to perform signal
processing tasks. Therefore, novel energy saving schemes taking these
often ignored sources of energy consumption into account are
required. We discuss the state-of-the-art and challenges to be
addressed in the development of 5G systems. Particular focus is given
to approaches that save energy by switching off hardware or by
adapting the transmission strategy. In more detail, if properly
exploited, spatial and temporal fluctuations of traffic in real
networks create opportunities to switch off hardware. We show machine
learning and statistical tools that can be used to predict the periods
of low traffic, and we also provide some observations on the evolution
of traffic patterns in current networks. These observations may be
used to guide the selection of prior knowledge to be incorporated into
traffic forecast tools for future networks. Even with good traffic
forecast tools, determining network elements that can be switched off
is not a trivial task because of the possible interference coupling
among active network elements. We therefore review general tools that
can be used to study a plethora of interference models in a unified
way. These tools are not restricted to the domain of wireless
communication, and we point out some simple and useful (and
perhaps novel) results that can greatly simplify the analysis of
coupled systems. The traffic and interference models are then used to
pose the (typically intractable) energy saving optimization problems,
and we present heuristics that i) have a strong analytical
justification and ii) have the potential to scale to problems of huge
dimensions, as required in dense networks.

\section*{\small ENERGY EFFICIENCY IN WIRELESS NETWORKS}

In the literature, there are different notions of energy efficiency, and selecting a suitable definition is a multi-faceted problem with profound theoretical and practical implications.

In \emph{communication systems with finite energy constraints,} it is
natural to relate the energy efficiency to the amount of energy that
we need to transmit a finite number of bits subject to a given error
probability. This fact has led researchers to consider the capacity
per unit energy (or the capacity of finite-energy channels in bits),
where the energy of the codewords is kept finite as the code length
tends to infinity \cite{Gallager88} (see also the discussion in
\cite[p.15]{Goldsmith02}). This notion, which measures the maximum
rate per unit energy \emph{(bit per second per Joule),} is difficult
to handle with the framework of classical multi-user information
theory. Therefore, information-theoretic studies have typically
considered the notion of energy per one bit \emph{(power divided by
  data rate),} which is defined as the amount of energy that is
required for reliable (i.e., asymptotically error-free) communication
of one bit of information at some rate \cite{Verdu02}. We emphasize
that \emph{the two notions are not equivalent} because, if the number
of bits tends to infinity at some rate, and the energy per bit is
fixed, then the total energy used for transmission tends to
infinity \cite{Goldsmith02}. In particular, the notion of energy per
bit, which commonly only considers the energy radiated by antenna, is
often used to show that recent techniques such as advanced multi-user
communication, massive MIMO, interference alignment, and network
coding are energy-efficient solutions. However, as the following
example shows, when other sources of energy consumption (such as
hardware and signal processing operations) are taken into account,
then the energy savings provided by these solutions may not be so
clear. For convenience, in the text that follows, we use the general
term \emph{radiated energy} to refer to the energy radiated by
antennas, and we reserve the term \emph{operating energy} to refer to the
remaining sources of energy
consumption. 

\begin{example}
  The study in \cite{ZemlianovEtAl2005} investigates the impact of
  \emph{infrastructure nodes (base stations, access points, etc.)} on
  the throughput scaling. Inspired by the operation of current
  practical systems, where the interference is treated as noise and
  the transmission is considered successful if a given
  signal-to-interference-plus-noise (SINR) ratio is attained, the
  authors propose a communication scheme for random networks (i.e.,
  wireless devices placed randomly in an uniform and independent
  manner, whereas infrastructure nodes are placed arbitrarily in a
  predefined manner, independent of the placement of the wireless
  devices) that achieves the throughput scaling of $\varTheta
  \bigl(\tfrac{m}{n}\bigr)$, provided that
  $m(n)\in\omega\bigl(\sqrt{\tfrac{n}{\log n}}\bigr)$ and $m(n)\in
  O\bigl(\tfrac{n}{\log n}\bigr)$. Here, $m$ denotes the number of
  base stations, and $n$ is the number of users. It can be shown
  \cite[Deliverable D4.2]{greennets} that, for this scheme and this
  scaling of base stations, by choosing $m(n) = (\tfrac{n}{\log n})^b$
  with $b \in (\tfrac{1}{2},1]$, {the
    radiated energy per information bit diminishes to zero as the
    number of users tends to infinity. In contrast, the operating
    energy consumed per information bit} $\text{E}_b(n)$
  increases at the order of $\text{E}_b(n)\in \Theta
    \bigl(n\bigl(\tfrac{\log{n}}{n}\bigr)^b \bigr)$.
\end{example}

 The above example reveals that, although the
transmit energy per bit may vanish as the number of nodes increases
{in the considered scenario}, the \staticEnergy per transmitted
information bit {increases
  without bound (in the best case as $\Theta(\log n)$)}. Therefore, in
highly dense networks ($m$ and $n$ large), the energy consumed by
hardware is dominant, so we draw the following important conclusion:

\begin{fact}
  Advanced multi-user communication strategies such as cooperative (relaying) techniques, massive MIMO,
  interference alignment, and network coding are highly promising
  approaches to push the performance of wireless networks with respect
  to throughput, delay, and error probability to orders of magnitude
  beyond the performance limits of contemporary cellular
  networks. These technologies can decrease the transmit
  energy per bit, but, alone, they do not reduce significantly the
  \staticEnergy.  On the contrary, they may lead to a significant
  increase of the \staticEnergy. This is due to the increased
  number of antennas (and the accompanying hardware), and the energy-expensive and time-consuming
  signal processing algorithms.
\end{fact}

One of the major design principles to reduce the consumption of the \dynamicEnergy and the \staticEnergy is to adjust the capacity of
the network to the demand. In particular, in delay-sensitive applications, one option is to devise load-dependent algorithms that deactivate network elements in a coordinated manner in order to provide the desired coverage and throughput performance at any given point in time. The energy saving capabilities of these algorithms is limited by a fundamental trade-off between energy efficiency and delay constraints, which, to a large extent, remains an open problem \cite{FundGreenTradeoffs11}. We can take this idea one step further and consider that the algorithms are also able to choose the most suitable multi-user communication strategy (e.g., network coding, MIMO technique, etc.) for the active network elements. Signal processing tools that can be used to develop and to support these load-adaptive schemes are the topic of the remaining sections. 

\section*{\small TRAFFIC PATTERNS}

In current communication networks, traffic typically follows a roughly periodic pattern; traffic is high during day and low at night, with some local variations depending on whether we consider industrial, commercial, residential, or rural areas. These spatial and temporal fluctuations create opportunities to save energy by switching off unnecessary network elements. To exploit this approach to its fullest potential, we need reliable forecasts of local traffic, a task that calls for machine learning algorithms and statistical tools.

In general, the prediction power of learning algorithms improves as we increase the amount of available prior information, and one of the most natural assumptions to use in traffic forecasts is the rough periodicity of the time series. However, especially in future networks, assumptions of this type have to be used carefully for two main reasons. First, in current networks much of the traffic is generated by users, so the patterns are correlated to those of the human activity. For instance, users are more likely to watch streaming videos during day or evening than very late at night. In contrast, in future networks, it is envisioned that machine-to-machine and device-to-device communications will be ubiquitous, and, for the devices requesting such services, the time of the day when data is sent may be irrelevant, or it may be even desirable to send data when other users are less active (e.g., noncritical software updates or backups could be performed while users are sleeping to avoid unnecessary congestion in the network). Second, current studies showing the coarse periodicity of traffic typically consider traffic aggregated over regions of the network, but at a local scale (individual cells of the network) the patterns many not necessarily follow the global trend too closely. This last observation has already been noticed in historical data coming from a real network of a large European city \cite[Deliverable D6.2]{greennets}, where the time series of key performance indicators (KPIs) related to data traffic are less regular than those related to voice calls. In Fig.~\ref{fig:loadEvol} we show synthetic voice and data traffic with similar statistical properties to those found in a typical cell of the above-mentioned real network. Note that, when data traffic is considered, bursts of traffic are frequently observed in periods when voice traffic is predictably low. The practical implication of these observations is that, especially at a local scale,  forecasting algorithms should take into account the type of the service. The clear periodicity of current traffic patterns may not be necessarily present in future systems. In particular, prediction algorithms for data traffic should use statistically robust methods because of the irregularity of the time series, whereas prediction algorithms for voice traffic may be able to safely assume the coarse periodicity observed in Fig.~\ref{fig:loadEvol}(a).

\begin{figure*}[ht]
 \centering
 \includegraphics[width=0.3\textwidth]{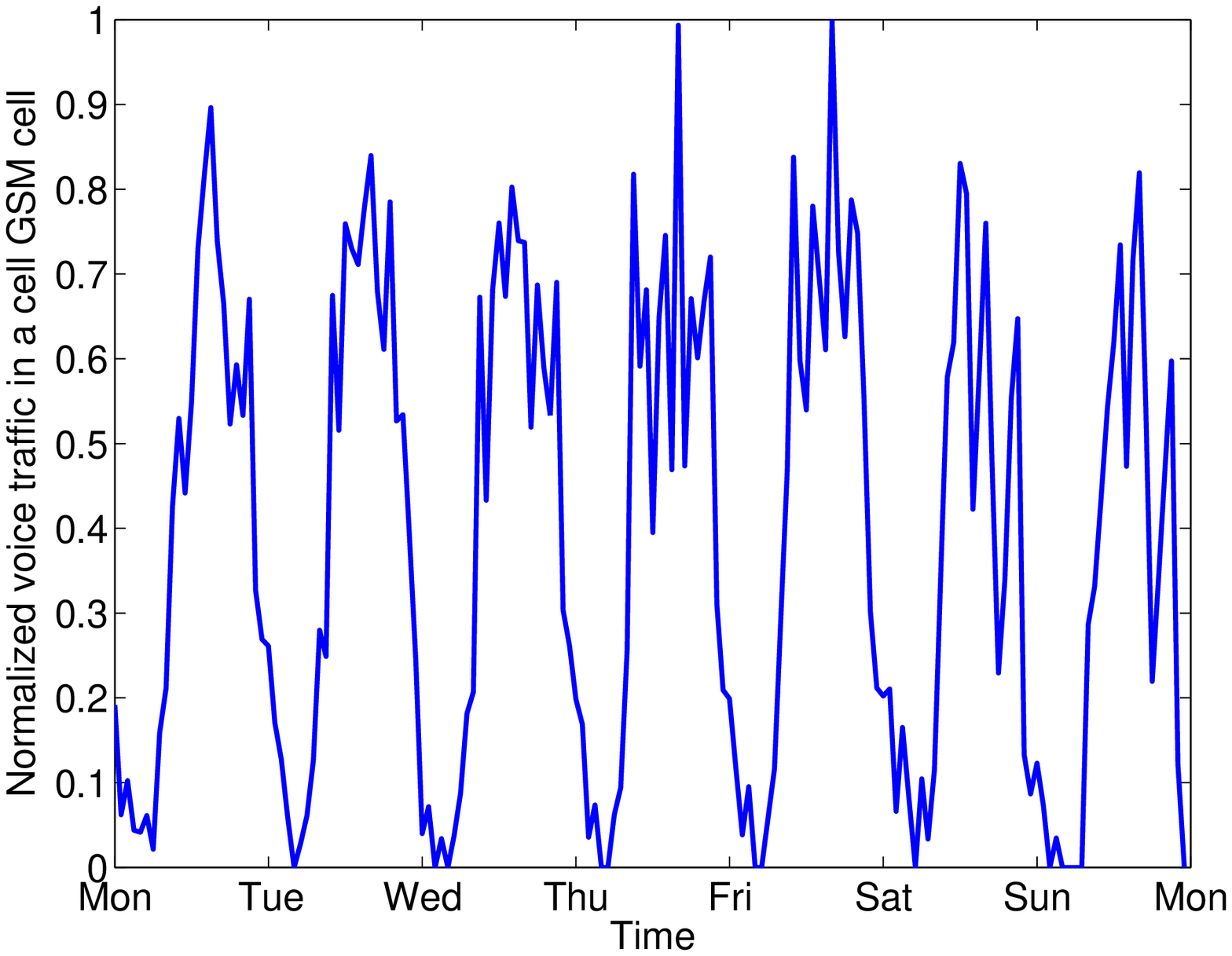}
 (a) 
 \includegraphics[width=0.3\textwidth]{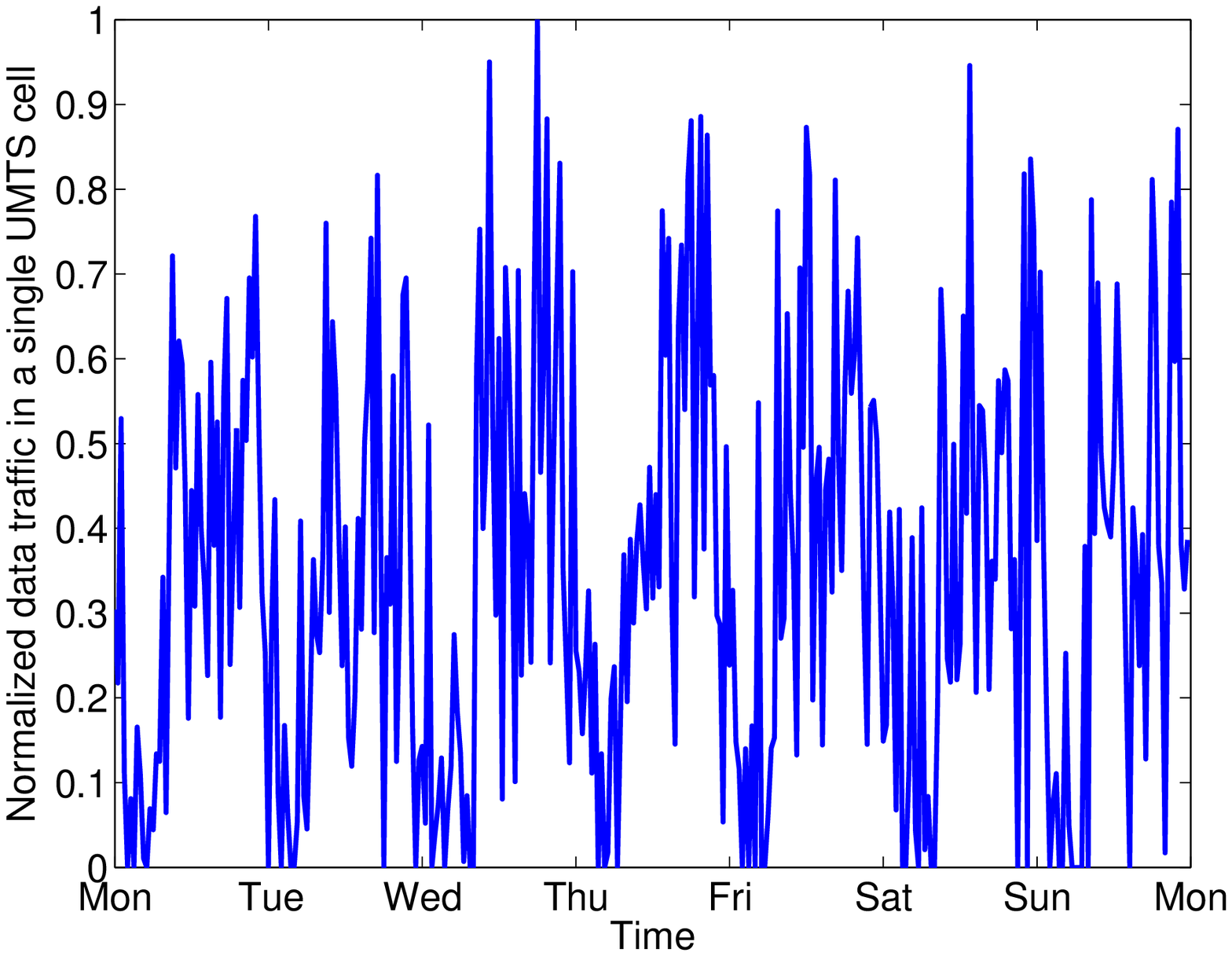}
 (b) 
 \includegraphics[width=0.3\textwidth]{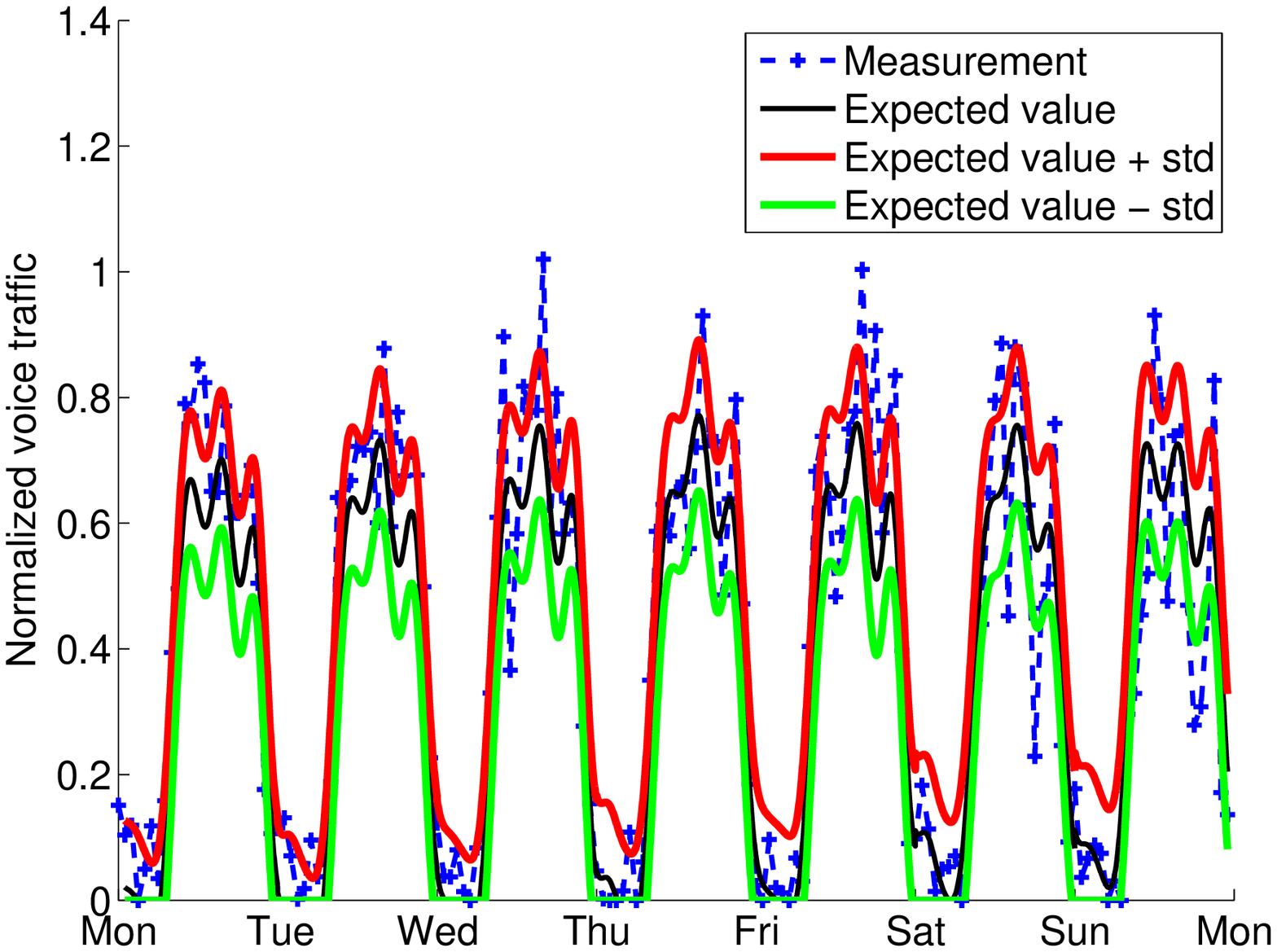}
(c)
 \caption{Normalized traffic as a function of time. (a) Synthetic voice traffic in a single cell of a global system for mobile communications (GSM) network. (b) Synthetic data traffic in a single cell of a universal mobile telecommunications system (UMTS) network. (c) Forecast for voice traffic obtained with the framework of Gaussian processes (``std'' stands for standard deviation) --- three weeks of data are used for training.} 
 \label{fig:loadEvol}
\end{figure*}

To produce forecasts of time series strongly related to voice calls, we can use approaches similar to those used for electrical load forecasting because electrical load in power grids and voice calls in cellular networks present similar periodicity. This knowledge and other contextual information, such as the presence of holidays and major events, can be easily incorporated into algorithms based on Gaussian processes (GPs) \cite{ras06} and other kernel-based methods \cite[Deliverable D4.1]{greennets}. The main idea is to choose a covariance or kernel function based on the observed features of the time series under consideration. In particular, the framework of GPs is a promising approach that enables us to specify those features merely in general terms, and confidence intervals for the predictions are readily available. For example, we assume that the time series in Fig.~\ref{fig:loadEvol}(a) has a clear periodic pattern, so we can specify a periodic covariance function and leave the period as a free parameter (hyperparameter) to be selected by maximizing a merit function with a very natural probabilistic interpretation \cite[Ch. 5]{ras06} (e.g., the marginal log-likelihood). This example is not particularly interesting because the period can be easily obtained by simply looking at the data, but the approach can be straightforwardly applied to capture more subtle features such as small variations of traffic according to the day of the week. To this end, we can design ``atomic'' covariance functions, each of which responsible for capturing one desired feature of the time series. Then we construct the final covariance function by combining the atomic functions with operations that preserve covariance functions, and we compute the hyperparameters by maximizing the marginal log-likelihood.  In Fig.~\ref{fig:loadEvol}(c) we show results for the prediction of voice traffic obtained by following the general guidelines for the selection of the covariance function outlined in \cite[Ch.~5.4.3]{ras06}. For this result, we use synthetic data, but the same algorithm provides good forecasts for most cells of a real network. However, if the time series contain too many bursts of traffic, such as that depicted in Fig.~\ref{fig:loadEvol}(b), the traditional framework of GPs may have bad generalization properties. In such cases, robust statistical tools are required, and we now review possible approaches.

It has been observed that, in the short/medium term (up to a couple of months), samples of traffic-related KPIs for either working days or holidays can be assumed to come from i.i.d. random variables for most cells of a real network \cite[Deliverable D6.2]{greennets} if they are spaced by multiples of 24h (indication of the validity of this assumption can be obtained with the turning point test \cite{le2010performance}). As a result, we can use simple robust tools based on order statistics, such as tolerance intervals \cite{le2010performance}, to obtain knowledge about upper bounds for traffic at any given hour of the day.  More precisely, let $X_{1:n}\le\ldots\le X_{n:n}$  be the sorted values of the i.i.d. random variables $X_1,\ldots,X_n$ corresponding to traffic measurements $x_1,\ldots,x_n$ for a given hour of the day. Denote the cumulative distribution function of the random variables by $F(x)$ and the inverse cumulative distribution function by $F^{-1}(p)=\sup\{x~|~F(x)\le p\}$, where $p\in~]0,1[$. Then, for a fixed quantile $p$, the following holds \cite{barry}: $ P\left(F(X_{k:n}) > p \right)=1-\sum_{i=k}^n\left(\begin{matrix}{n}\\{i}\end{matrix}\right) p^i~(1-p)^{n-i}$, which is an exact value that does not depend on the distribution of the random variables. If this probability is sufficiently low and $p$ sufficiently high, we can configure the cell or base station to serve at most traffic value $X_{k:n}$. Similar strikingly simple and exact results exist to other types of intervals (e.g., prediction intervals), and good agreement between the theoretical and empirical results has been obtained with real network data \cite[Deliverable D6.2]{greennets}.

A major limitation of the above robust approaches is that they are unable to detect trends and to capture correlations between samples from consecutive hours of the day (samples from different hours are analyzed independently). Correlations and trends can be captured by robust machine learning tools such as those described in \cite{takeuchi2006}. To  improve  further the estimates, we can also try to exploit temporal and spatial correlations among  cells. Extensions of this type should consider carefully the computational complexity because of the large number of network elements in future networks. Unfortunately, obtaining data sets for research from real networks is difficult, a fact that has limited the literature on this important topic. 

\section*{\small NETWORK INTERFERENCE CALCULUS}
Even with good traffic forecasts, identifying the best action to save energy still remains a difficult problem because of the possible interference coupling among active network elements. Therefore, it comes as no surprise that energy saving algorithms also need at least a rough estimate of the interference patterns, and we now turn the attention to some basic results on interference calculus \cite{yates95,martin11}, a general mathematical framework that unifies many interference models in wireless systems. The presentation is heavily based on the study in \cite{yates95}, which shows algorithms for power-control in CDMA systems. From a mathematical perspective, these algorithms solve general fixed point problems, so they have been used in applications different from that originally envisioned. We expect these algorithms based on interference calculus to play an important role in the analysis of future systems owing to the generality of the framework.

\subsection*{\small Standard Interference Functions}

In the following discussion $\real_+$ denotes the set of non-negative reals, $\real_{++}$ is the set of strictly positive reals, and $\signal{1}$ is a vector of ones. Inequalities involving vectors should be understood as element-wise inequalities. Interference functions are defined as follows (for convenience, our definitions are slightly different from that originally stated in \cite{yates95}):

\begin{definition}
\label{definition.inter_func}
 A function $I: \real_+^M \to \real_{++}$ is said to be a standard interference function if the following axioms hold:\par 
 1. ({\it Scalability}) $\alpha {I}(\signal{x})>I(\alpha\signal{x})$ for all $\signal{x}\in\real^M_+$ and all $\alpha>1$. \par
 2. ({\it Monotonicity}) ${I}(\signal{x}_1)\ge I(\signal{x}_2)$ if $\signal{x}_1\ge \signal{x}_2$. \par 
Given $M$ standard interference functions $I_i:\real^M_+\to\real_{++}$, $i=1,\ldots,M$, we call the mapping $\mathcal{J}:\real^M_+\to\real_{++}^M$ with $\mathcal{J}(\signal{x}):=[I_1(\signal{x}),\ldots, I_M(\signal{x})]^T$ a ``standard interference mapping'' or simply ``interference mapping.''

\end{definition}

Checking whether a given function is a standard interference function by using the definition is not necessarily easy. Fortunately, the following proposition shows that a large class of functions frequently used to model wireless systems are standard interference functions. We note that the simple result shown below has been explicitly mentioned in the nonpublic report \cite[D52]{greennets} (see also \cite{feh2013} and the references therein).

\begin{proposition}
\label{proposition.pos_conc}
 Concave functions $I:\real^M_{+}\to\real_{++}$ are standard interference functions.
\end{proposition}

The usefulness of Proposition~\ref{proposition.pos_conc} lies in the fact that there exist many simple and well-known techniques to identify concave and convex functions \cite{boyd}. The converse of Proposition~\ref{proposition.pos_conc} does not hold in general. However, we can expand the class of functions that can be easily identified with the help of Proposition~\ref{proposition.pos_conc} by using the following operations that preserve interference functions/mappings. 

\begin{fact} (Operations that preserve standard interference functions/mappings \cite{yates95,martin11}:) \par
 \label{fact.op_prev}
 1. Standard interference functions are closed under finite addition and multiplication by strictly positive constants. For instance, if $I_1:\real^M_{+}\rightarrow\real_{++}$ and $I_2:\real^M_{+}\rightarrow\real_{++}$ are standard interference functions, then $I^{\prime}(\signal{x}) = \alpha_1 I_1(\signal{x})+\alpha_2 I_2(\signal{x})$ for $\alpha_1,\alpha_2>0$ is a standard interference function.\par
 2. If ${I}_i:\real^M_{+}\rightarrow\real_{++}$ ($i\in\{1,\ldots,N\}$, $N\in\Natural$) are standard interference functions, then ${I}^\prime(\signal{x}):=\min_{i\in\{1,\ldots,N\}}{I}_i(\signal{x})$ and ${I}^{\prime\prime}(\signal{x}):=\max_{i\in\{1,\ldots,N\}}{I}_i(\signal{x})$ are standard interference functions. \par
 3. Standard interference mappings are closed under finite composition. For example,  if $\mathcal{J}_1:\real^M_{+}\rightarrow\real^M_{++}$ and $\mathcal{J}_2:\real^M_{+}\rightarrow\real^M_{++}$ are standard interference mappings, then $\mathcal{J}^{\prime}(\signal{x})=\mathcal{J}_1(\mathcal{J}_2(\signal{x}))$ is a standard interference mapping.\par 
 
\end{fact}

We are mostly interested in studying fixed points of standard interference mappings, and the following result can be used for this purpose. When interference calculus is used to investigate the performance of communication systems, the fixed points describe, for example, the load or interference experienced by network elements. 

\begin{fact}
 \label{fact.int_func} (Selected properties of standard interference mappings \cite{yates95}:) 
Let $I_i:\real^M_+\to\real_{++}$ be a standard interference function for every $i\in\{1,\ldots,M\}$, and consider the corresponding mapping $\mathcal{J}:\real^M_+\to\real_{++}^M$  given by $\mathcal{J}(\signal{x}):=[I_1(\signal{x}),\ldots, I_M(\signal{x})]^T$. Then the following holds:
\begin{enumerate}
\item If the mapping $\mathcal{J}$ has a fixed point (i.e., $\emptyset\ne\mathrm{Fix}(\mathcal{J}):=\{\signal{x} \in \real^M_{+} ~|~ \signal{x}=\mathcal{J}(\signal{x})\}$), then the fixed point is unique. 
\item The mapping $\mathcal{J}$ has a fixed point if and only if there exists $\signal{x}^\prime \in\real^M_{++}$ satisfying $\mathcal{J}(\signal{x}^\prime)\le \signal{x}^\prime$. 
\item If $\mathcal{J}$ has a fixed point, then the sequence $\{\signal{x}_{n}\}_{n\in\Natural}$ generated by $\signal{x}_{n+1}=\mathcal{J}(\signal{x}_{n})$ satisfies the following: \par
 i) For an arbitrary vector $\signal{x}_{0}\in\real^M_+$, the sequence $\{\signal{x}_n\}_{n\in\Natural}$ converges to the fixed point $\signal{x}^\star\in\mathrm{Fix}(\mathcal{J}$). \par 
 
 ii) If $\signal{x}_{0} = \signal{0}$, then the sequence $\{\signal{x}_n\}_{n\in\Natural}$ is monotonously increasing; i.e., $\signal{x}_{n+1}\ge \signal{x}_n$. \par
 
 iii) If $\mathcal{J}(\signal{x}_0)\le \signal{x}_0$, then the sequence $\{\signal{x}_n\}_{n\in\Natural}$ is monotonously decreasing; i.e., $\signal{x}_{n+1}\le \signal{x}_{n}$.

\end{enumerate}
\end{fact}

\begin{remark}
\label{remark.nonheuristic}
Suppose that, in addition to being an interference mapping, $\mathcal{J}:\real^M_+\to\real_{++}^M$ is also upper bounded; i.e., $\mathcal{J}(\signal{x})\le B~\signal{1}$ for some fixed $B\in\real$ and every $\signal{x}\in\real^M_+$. In this case, Facts.~\ref{fact.int_func}.1 and \ref{fact.int_func}.2 guarantee the existence of a unique fixed point $\signal{x}^\star$, and Fact~\ref{fact.int_func}.3 suggests the following iterative procedure to compute the fixed point.  Produce in parallel two sequences $\{\signal{x}_{n+1}^\prime=\mathcal{J}(\signal{x}_n^\prime)\}$ and $\{\signal{x}_{n+1}^{\prime\prime}=\mathcal{J}(\signal{x}_n^{\prime\prime})\}$, where $\signal{x}_0^\prime=\signal{0}$ and $\signal{x}_0^{\prime\prime}=B ~\signal{1}$. Fact~\ref{fact.int_func} shows that $\signal{x}_{n}^\prime \le \signal{x}^\star \le \signal{x}_{n}^{\prime\prime}$ for every $n\in\Natural$ and that both sequences $\{\signal{x}_n^\prime\}$ and $\{\signal{x}_n^{\prime\prime}\}$ converge to $\signal{x}^\star$. Furthermore, $\signal{x}^{\prime}_{n+1}\ge \signal{x}^{\prime}_{n}$ and $\signal{x}^{\prime\prime}_{n+1}\le \signal{x}^{\prime\prime}_{n}$.  We can therefore stop the algorithm at iteration $n$ whenever $\|\signal{x}_{n}^\prime - \signal{x}_{n}^{\prime\prime}\|_\infty \le \epsilon$ is satisfied, where $\epsilon>0$ is the desired precision. The algorithm is guaranteed to terminate with a finite number of iterations. In other words, $\signal{x}_{n}^\prime$ and $\signal{x}_{n}^{\prime\prime}$ can serve as lower and upper bounds, respectively, for $\signal{x}^\star$. 
\end{remark}

We now relate these results to load estimation in multi-carrier systems with fast link adaptation (as envisioned in 5G systems). The couplings of the models shown below have been originally studied on a case-by-case basis, but recently their connection to interference calculus has been independently established in \cite[Deliverable D5.2]{greennets}\cite{feh2013}. 

\subsection*{\small Load estimation in wireless networks}
The model described below is based on the discussion in \cite{siomina12}, and this or similar models have been used for various network optimization tasks for many years. Later we use this model to gain insight onto the challenges associated with the formulation of energy saving optimization problems.

In more detail, we focus on a cellular radio network with $M$ base stations (or cells) and $N$ test points (an abstract concept to represent demand of users in a given region). We denote the set of base stations by $\setm = 
\left\{ 1,2,...,M \right\}$ and the set of test points by $\setn = \left\{1,2,...,N\right\}$. The quality-of-service (QoS) requirement of each test point $j\in\setn$ is represented by a minimum amount of data $d_j\in\real_{++}$ that needs to be sent during a unit of time. We denote by $\signal{X} \in\{0,1\}^{M\times N}$ the assignment matrix; the component of its $i$th row and $j$th column takes the value $x_{i,j}=1$ if test point $j$ is assigned to base station $i$ or the value $x_{i,j}=0$ otherwise (we also assume that each base station serves at least one test point). The power gain between base station $i$ and test point $j$ is denoted by $g_{i,j}\in \real_+$. Each base station $i\in\setm$ transmits with fixed power spectral density per minimum resource unit (e.g., resource blocks in multi-carrier systems, time slots in TDMA systems, etc.) in scheduling, which we denote by $P_i\in\real_{++}$. The vector of load of the base stations is given by $\signal{\rho}:=[\rho_1,\ldots,\rho_M]^T \in \real_+^M$, where $\rho_i\in\real_{+}$ is the load at base station $i$. In words, the load $\rho_i$ is defined as the ratio between the number of resource units requested by test points served by base station $i\in\setm$ and the number $K$ of resource units available in the system. We can obtain an estimate of $\signal{\rho}$ by solving the following system of nonlinear equations \cite{siomina12}:
\begin{align}
\label{eq.system}
 \begin{matrix}
 \rho_1=I_1(\signal{\rho}), &  \cdots, & \rho_M=I_M(\signal{\rho}),
 \end{matrix}
\end{align}
where 
\begin{align}
\label{eq.loadi}
I_i(\signal{\rho}):=\sum_{j\in\setn} \dfrac{d_j~x_{i,j}}{K \omega_{i,j}(\signal{\rho})},
\end{align}
$\sigma^2$ is the noise power, and 
\begin{align}
\label{eq.omega}
 \omega_{i,j}(\signal{\rho}) := {B \log_2 \left(1+{\dfrac{P_i~ g_{i,j}}{\eta\left(\sum_{l\in\setm\backslash\{i\}} P_l~ g_{l,j} ~ \rho_l + \sigma^2\right)}}\right)}
\end{align}
is the the spectral efficiency (i.e., the effective bit rate per resource unit) of the link connecting base station $i$ to test point $j$. In \refeq{eq.omega}, $\omega_{i,j}$ depends on the effective bandwidth per resource unit $B$ and the SINR scaling factor $\eta$, which are parameters that are typically fitted from simulations of measurements of the actual spectral efficiency $\omega_{i,j}$ as a function of $\signal{\rho}$ for a particular network configuration (choice of schedulers, system bandwidth, MIMO transmission scheme, etc.). We refer the reader to \cite{mogensen07} for additional details. Intuitively, each term in the sum in \refeq{eq.loadi} is the fraction of resource units, relative to the total number of resource units K, that the test point $j$ requests from base station $i$ if data rate $d_j$ is desired. 

A positive  vector $\signal{\rho}^\star=[\rho_1^\star,\ldots,\rho_M^\star]^T$ is a solution of the system of nonlinear equations in \refeq{eq.system} if and only if $\signal{\rho}^\star$ is a fixed point of $\mathcal{J}(\signal{\rho}):=[I_1(\signal{\rho}),\ldots,I_M(\signal{\rho})]^T$ (i.e., $\signal{\rho}^\star\in\mathrm{Fix}(\mathcal{J})$). Once a solution is obtained (assuming that it exists and is unique) we can verify whether the network configuration can support the traffic demand by checking whether $\rho^\star_i\le 1$ for every $i\in\{1,\ldots,M\}$; i.e., base stations do not use more resource units than available, in which case we say that the network configuration is feasible. As a result, answering questions regarding uniqueness and existence of a fixed point of $\mathcal{J}$ (and also iterative methods to compute fixed points) is crucial to verify feasibility of networks. To this end, we can use the result in Proposition \ref{proposition.pos_conc} and Fact~\ref{fact.int_func}. More precisely, we note that $I_i(\signal{\rho})$ is a standard interference function because it is positive  and concave  (see Proposition \ref{proposition.pos_conc}). Concavity of $I_i$ follows from simple facts \cite{boyd}: i) $f(x):=1/\log_2(1+1/x)$ is a concave function on the domain $\real_{++}$, ii) composition of concave functions with affine transformations preserves concavity, and iii) the set of concave functions is closed under addition and multiplication by strictly positive real numbers. (See \cite{siomina12} for an alternative argument.) By identifying $\mathcal{J}$ as a standard interference mapping, we can now use the known results in Fact~\ref{fact.int_func} to compute fixed points, answer questions regarding existence and uniqueness of fixed points, etc. Note that some previous studies have not identified the functions $I_i$ as standard interference functions. The advantage of working with the framework of interference calculus (in addition to showing that many existing results are direct consequences of Fact~\ref{fact.int_func}) becomes clear when we extend the interference coupling model in \refeq{eq.system} by using simple operations that preserve interference mappings. By doing so, we can analyze more complex communication systems than those described above (see also \cite{feh2013}\cite[Deliverable D5.2]{greennets} for more details on the examples).

\begin{example}
\label{example.upper_bound}
 Assume that the spectral efficiency $\omega_{i,j}$ in \refeq{eq.omega} is upper bounded by $\bar{\omega}$, which is a natural assumption in real systems. In such cases, we can evaluate the feasibility of a network by computing the fixed point of the mapping $\mathcal{J}^\prime(\signal{\rho}):=[I_1^\prime(\signal{\rho}), \ldots, I_M^\prime(\signal{\rho})]$, where 
$I_i^\prime(\signal{\rho}):=\sum_{j\in\setn} \max\left\{\dfrac{d_j~x_{i,j}}{K \omega_{i,j}(\signal{\rho})},~\dfrac{d_j~x_{i,j}}{K \bar{\omega}}\right\}
$. Note that we can easily verify that $I_i^\prime$ is a standard interference function by using Fact~\ref{fact.op_prev}.
\end{example}

\begin{example}
 Suppose that the system considered in Example \ref{example.upper_bound} is not feasible; i.e., the interference mapping $\mathcal{J}^\prime$ either does not have a fixed point or, if a fixed point exists, some of its components have value strictly larger than one (in which case at least one of the base stations require more resource units than available in the system). In such cases, the system may still work with overloaded base stations dropping users (i.e., only a fraction of the traffic of the overloaded base stations is served), but the fixed point of $\mathcal{J}^\prime$ is not useful to indicate the load in non-overloaded base stations. To capture this feature of real systems, we can impose limits on the maximum possible load by using the interference mapping $\mathcal{J}^{\prime\prime}(\signal{\rho}):=[I_1^{\prime\prime}(\signal{\rho}), \ldots, I_M^{\prime\prime}(\signal{\rho})]$, where $I_i^{\prime\prime}(\signal{\rho}):=\min\left\{I_i^\prime(\signal{\rho}),~ 1\right\}$. We have already seen that each function $I_i^\prime$ is a standard interference function, so Fact~\ref{fact.op_prev} shows that $\mathcal{J}^{\prime\prime}$ is also a standard interference function. By noticing that $\mathcal{J}^{\prime\prime}(\signal{\rho}) \le \signal{1}$ for every $\signal{\rho}\in\real_{+}^M$, Fact~\ref{fact.int_func}.2 shows that $\mathcal{J}^{\prime\prime}$ is guaranteed to have a fixed point, which can be computed with the scheme with the non-heuristic stopping criterion in Remark~\ref{remark.nonheuristic}.  
\end{example}

The aforementioned interference coupling models are also useful to highlight limitations of interference calculus, which should be addressed in future extensions of the framework. As discussed in the introduction, future systems will be composed of combinations of many energy efficiency transmission schemes. In particular, when network elements are cooperative or apply interference-exploiting methods, extending the above models while remaining under the framework of interference calculus is difficult. For example, although the model in \refeq{eq.omega} can be adjusted to account for some limited advanced communication strategies (presence of intelligent schedulers, MIMO transmitters, etc.), it is an approximation that can be too crude in future systems where cooperation among network elements will be taken to completely new levels. We refer the reader to \cite{martin11} for recent advances in the field.

\section*{\small ALGORITHMS FOR ENERGY SAVINGS}

We now turn the attention to algorithms that have the objective of switching off as many network elements as possible while satisfying constraints such as coverage and data rate requirements. These algorithms typically use as an input the traffic forecasts and interference models described in previous sections. To avoid notational clutter, we assume for the moment that all network elements consume the same amount of energy and that the \dynamicEnergy is negligible. In particular, the latter assumption is an acceptable approximation in current cellular systems \cite{correia10}. All assumptions are later dropped to take into account heterogeneous systems with  hardware more energy efficient than that available today.

We associate network elements of a communication system with a vector $\signal{x}=[x_1,\ldots,x_M]^T\in\{0,1\}^M$, where $M$ is the number of network elements and each variable $x_i\in\{0,1\}$ takes the value one if network element $i$ is active, or the value zero otherwise. By $\mathcal{X}$ we denote the set of configurations $\signal{x}\in\{0,1\}^M$ satisfying some required constraints (e.g., capacity constraints). The mathematical problem that energy-saving algorithms try to solve can be typically described as a variation of the following combinatorial problem:
\begin{align}
 \mathrm{minimize} &\quad \|\signal{x}\|_0 \label{eq.l0norm} \\
 \mathrm{subject~to} & \quad \signal{x}\in\mathcal{X}, \nonumber \\
 & \signal{x}\in\{0,1\}^M \label{eq.discrete_general}
\end{align}
where $\|\signal{x}\|_0$ is the function, informally called $l_0$-norm ($\|\cdot\|_0$ does not satisfy all axioms of norms), which returns the number of nonzero elements of the vector $\signal{x}$. The above energy problems are typically NP-hard, so we cannot expect to solve them both fast and optimally, especially when considering the densification of future networks, which will lead to problems of huge dimensions. If optimality is desired, we can use a branch and bound algorithm, but solving even fairly small problems often takes a very long time \cite{joshi09}. As a result, with the densification of the networks, recent studies \cite{pollakis12,su13,Niu2010CellZooming,yamada11} have focused on fast, but suboptimal heuristics, and many of them \cite{pollakis12, joshi09, su13,yamada11} build upon theoretically sound methods that aim at solving discrete optimization problems by using the $l_1$-norm as a proxy of the $l_0$-norm. The reason for this approximation is that the $l_1$-norm is a (convex) sparsity promoting norm \cite{candes08b,boyd,joshi09}, and convex problems are typically easier to solve than nonconvex problems. (We can also interpret the $l_1$-norm as the convex envelope of the $l_0$-norm for an appropriate domain \cite{boyd}.) By also relaxing nonconvex constraints to convex constraints, many efficient optimization techniques become available \cite{boyd,yamada11}. In particular, we often replace the discrete constraint in \refeq{eq.discrete_general} by 
\begin{align} 
\label{eq.continuous}
\signal{x}\in[0,1]^M.
\end{align}
The solution of the resulting convex optimization problem (assuming that $\mathcal{X}$ is also a convex set) can then guide other heuristics to make the final hard decisions on the active set of network elements. This approach is used in, for example, the study in \cite{yamada11}, which considers an antenna selection problem in energy limited point-to-point communication systems with constraints given in terms of the required channel capacity. \par 

In the compressive sensing community, problems with the $l_0$-norm in the objective are increasingly being solved with a method called reweighted $l_1$-norm \cite{candes08b}, which in recent years are finding applications in cellular communication systems \cite{pollakis12,su13}. We first review the majorization-minimization (MM) algorithm in order to then explain these ideas.

{\small\bf The MM algorithm \cite{candes08b,sri11}:}  The discussion here follows closely that in the studies in \cite{candes08b,sri11}. Suppose that the objective is to minimize a function $f:\mathcal{X}\rightarrow \real$, where $\mathcal{X}\subset\real^M$. Unless the optimization problem has a very special structure that can be exploited, such as convexity, finding an optimal solution $\signal{x}^\star$ (provided that one exists) is intractable in general. To devise a suboptimal approach, assume that we are able to construct a function $g:\mathcal{X}\times\mathcal{X}\rightarrow \real$, hereafter called majorizing function, satisfying the following properties:
\begin{align}
 \label{eq.mmp1}
 f(\signal{x})\le g(\signal{x},\signal{y}),\quad\forall\signal{x},\signal{y}\in\mathcal{X}
\end{align}
\noindent and
\begin{align}
 \label{eq.mmp2}
 f(\signal{x})=g(\signal{x},\signal{x}),\quad\forall\signal{x}\in\mathcal{X}.
\end{align}
Then, starting from $\signal{x}_0\in\mathcal{X}$, the MM algorithm produces a sequence $\{\signal{x}_n\}\subset\mathcal{X}$ by
\begin{align}
\label{eq.mmit}
 \signal{x}_{n+1}\in\arg\min_{\signal{x}\in\mathcal{X}}g(\signal{x},\signal{x}_n).
\end{align}

We can verify that the sequence $\{f(\signal{x}_n)\}$ is monotonously decreasing: $f(\signal{x}_{n+1}) = g(\signal{x}_{n+1},\signal{x}_{n+1}) \le g(\signal{x}_{n+1},\signal{x}_{n}) \le g(\signal{x}_{n},\signal{x}_{n}) = f(\signal{x}_{n})$,
where the equalities follow from \refeq{eq.mmp2}, and the first and second inequalities follow from \refeq{eq.mmp1} and \refeq{eq.mmit}, respectively. As a result, $f(\signal{x}_n)\to c\in\real$ for some $c\ge f(\signal{x}^\star)$ as $n\to\infty$, where we assume that $\signal{x}^\star$ is a solution of the original optimization problem. In practice, we stop the algorithm when we observe no progress in the objective value, and we note that convergence of $\{f(\signal{x}_n)\}$ does not in general imply the convergence of the sequence $\{\signal{x}_n\}$. \par 

 The main challenge in applying the MM algorithm is to find a majorizing function $g$ such that the iteration in \refeq{eq.mmit} can be implemented efficiently. Fortunately, in some special cases of practical interest, we can construct    a majorizing function easily. For example, if $f$ can be decomposed as $f(\signal{x})=f_1(\signal{x})+f_2(\signal{x})$, where $f_1$ is a differentiable concave function and $f_2$ is a convex function, then we can use 
\begin{align}
\label{eq.majorizing}
g(\signal{x},\signal{y}) = f_1(\signal{y})+\nabla f_1(\signal{y})^T(\signal{x}-\signal{y})+f_2(\signal{x})
\end{align}
 as the majorizing function, where $\nabla f_1(\signal{y})$ is the gradient of $f_1$ at $\signal{y}$ (if $f_1$ is not differentiable, we can replace the gradient by an arbitrary subgradient). With this choice, the optimization problem in \refeq{eq.mmit} becomes a convex optimization problem provided that the set $\mathcal{X}$ is convex, hence it can be solved with efficient methods \cite{boyd,yamada11}. We can verify the validity of the property in \refeq{eq.mmp1} from the first-order characterization of concave functions \cite{boyd}. \par 

Let us turn the attention to the optimization problem in \refeq{eq.l0norm}-\refeq{eq.discrete_general} with the constraint in \refeq{eq.discrete_general} replaced by that in \refeq{eq.continuous}. It is well know that the $l_0$-norm satisfies the following \cite{candes08b,sri11}:
 \begin{align}
\label{eq.x0}
  \|\signal{x}\|_0 = \lim_{\epsilon\to 0}\sum_{i=1}^M \dfrac{\log(\epsilon+|x_i|)-\log(\epsilon) }{\log(1+\epsilon^{-1})},
 \end{align}
which, if each component $x_i$ of the vector $\signal{x}=[x_1,\ldots,x_M]^T$ is constrained to be non-negative as in \refeq{eq.continuous}, suggests the use of the function $f_\epsilon(\signal{x}) = \sum_{i=1}^M(\log(\epsilon+x_i)-\log(\epsilon))/\log(1+\epsilon^{-1})$ with a small design parameter $\epsilon>0$ as a approximation of the $l_0$-norm (other choices are possible \cite{candes08b}). Furthermore, note that the function $f_\epsilon(\signal{\signal{x}})$ is concave, so the MM algorithm can be used to generate a sequence of vectors $\{\signal{x}_n\}$ with decreasing objective value as discussed above. 

\subsection*{\small Energy saving optimization problems in cellular networks}
We now exemplify how to apply the above ideas to energy saving problems in cellular networks. (These problems also highlight the need to study interference coupling in energy efficient networks.) To this end, consider the notation introduced in the previous section on interference calculus, and assume that, for each test point $j\in\setn$, we have an estimate of the traffic demand per unit of time, which, as before, we denote by $d_j>0$. This estimate can be obtained with learning algorithms. We assume that the \staticEnergy of each base station (which here can also designate a cell, a radio unit, etc.) $i\in\setm$ is given by $c_i>0$, and the energy related to radiation is approximated by a function $f_i: [0,1]\to\real_+$ of the load $\rho_i\in[0,1]$. To detect base stations that can be switched off, we may try to solve the following problem:
\begin{align}
{\mathrm{min. }} & \sum_{i = 1}^M c_i |\rho_i|_0 + \sum_{i = 1}^M f_i(\rho_i)&  
 \label{eq.nonconvex}\\
\mathrm{s.t. }\quad & \rho_i = \sum^N_{j = 1} \frac{d_j}{K {\omega}_{i,j}(\signal{\rho})}x_{i,j}, & i \in \setm \label{eq.rho} \\
& \sum^M_{i = 1} x_{i,j} = 1 &  j   \in \setn \nonumber \\
& x_{i,j} \in \left\{ 0,1 \right\} &  i \in \setm,  j\in \setn \label{eq.discrete2} \\
& \rho_i\in [0,1] & i\in\setm \label{eq.last_const}
\end{align}
\noindent
where $\{x_{i,j}\}_{i\in\setm,j\in\setn}$ (assignment variables) and $\{\rho_i\}_{i\in\setm}$  (load at base stations) are the optimization variables, and $|x|_0$ is the function that takes the value 0 if $x=0$ or the value 1 otherwise. By solving problem \refeq{eq.nonconvex}-\refeq{eq.last_const}, base stations with no load (i.e., those with $\rho_i=0$) can be deactivated. In the problem, we assume that the constraints are feasible. If not, we can add slack variables and an additional $l_0$ or $l_1$ penalty norm to the merit function, but, for brevity, we do not consider such extensions here. One of the main complications to the above problem is the interference coupling appearing in the nonconvex constraint in \refeq{eq.rho}. We have already seen in previous sections that computing interference or load, even with fixed network configurations, is a nontrivial task. Therefore, it is common in the literature to fix the value of the spectral efficiency of the link connecting base station $i$ to test point $j$ to some constant $\tilde{\omega}_{i,j}$, which is obtained by considering the worst-case interference or an average case. Unfortunately, even with this simplification, the resulting problem is difficult to solve because it is a generalization of the standard bin-packing problem, which is NP-hard. To obtain a fast heuristic, we can use the MM algorithm as follows. First, as discussed in the previous section, we relax the discrete constraint in \refeq{eq.discrete2} to consider continuous values $x_{i,j} \in \left[ 0,1 \right]$. (In particular, this modification is natural when base stations are allowed to serve only a fraction of the traffic requested at test points.) Then, by also using the approximation discussed below \refeq{eq.x0}, we obtain the following optimization problem, which can be efficiently addressed with the majorization-minimization algorithm if $f_i(\rho_i)$ is a convex or concave function (see the discussion above \refeq{eq.majorizing}), and we note that a linear function is often considered a good approximation of the energy related to radiation \cite{correia10}: 

\begin{align}
{\mathrm{min. }} & \sum_{i = 1}^M c_i \dfrac{\log(\epsilon+\rho_i)-\log(\epsilon) }{\log(1+\epsilon^{-1})} + \sum_{i = 1}^M f_i(\rho_i)& \label{eq.last_problem}\\
\mathrm{s.t. } \quad & 0\le \rho_i = \sum^N_{j = 1} \frac{d_j}{K \widetilde{\omega}_{i,j}}x_{i,j} \le 1, \quad i \in \setm  \label{eq.worst_case} \\
& \sum^M_{i = 1} x_{i,j} = 1, \quad  j   \in \setn \nonumber \\
& x_{i,j} \in [ 0,1 ] \quad  i \in \setm,  j\in \setn  \label{eq.last_const2}
\end{align}

A solution of this modified problem provides a good indication of which base stations to switch off ($\rho_i$ close to zero), a result that can guide other heuristics to make the final hard decisions. Empirical evidence suggests that these algorithms based on convex optimization are competitive, in terms of energy savings, against heuristics that deal with the nonconvex problems directly. We also note that convex optimization techniques can naturally consider the energy related to radiation, and they can exploit the rich structure of the problems to decrease the computational effort (e.g., many assignments are impossible owing to the distances involved). We refer the reader to \cite{pollakis12} for a comparison of the above MM-based method against the cell zooming approach in \cite{Niu2010CellZooming}. In Fig.~\ref{fig:comparison} we compare the solution obtained with 10 iterations of the MM algorithm against the optimal solution of the problem in \refeq{eq.nonconvex}-\refeq{eq.last_const} with the constraints in \refeq{eq.rho} replaced by the worst-case constraints in \refeq{eq.worst_case}. To obtain discrete values for each assignment $x_{i,j}$ after applying the MM algorithm to the problem in \refeq{eq.last_problem}-\refeq{eq.last_const2}, we use the heuristic outlined in \cite{pollakis12}. All optimization problems have been solved with CPLEX, a standard commercial solver. The simulated network mimics an LTE system with 100 base stations where the \dynamicEnergy is neglected, which is done to simplify the process of obtaining optimal solutions. The bandwidth of each base station is set to 100 MHz, and the data rate requirement of each test point/user is $128$ kbps. All other parameters are exactly the same as the network simulated in \cite{pollakis12}. We observe in Fig.~\ref{fig:comparison} that, although we obviously loose optimality by applying the MM algorithm, the time to obtain a network with fairly low energy consumption grows slowly as we increase the number of test points, which stands in sharp contrast to the algorithm that obtains optimal solutions.  For a variation of the above ideas to the problem of energy efficiency of small cell networks with best effort users, we refer the reader to \cite{su13}. These results indicate that techniques based on convex optimization (which, unlike discrete heuristics, can easily deal with the energy related to radiation) have great potential to scale to very large problems. 

\begin{figure*}[ht]
 \centering
 \includegraphics[width=0.3\textwidth]{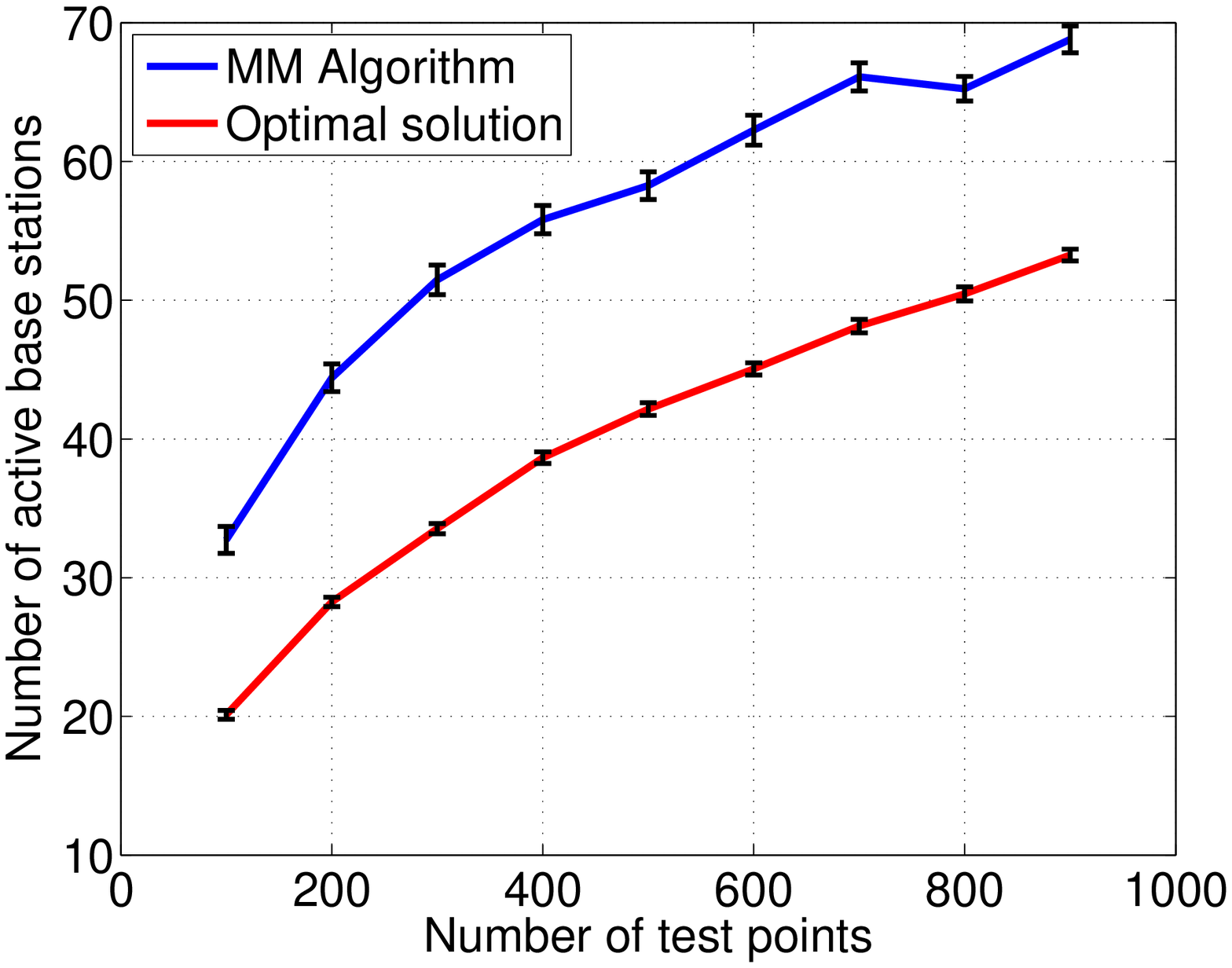}
 (a) 
 \includegraphics[width=0.3\textwidth]{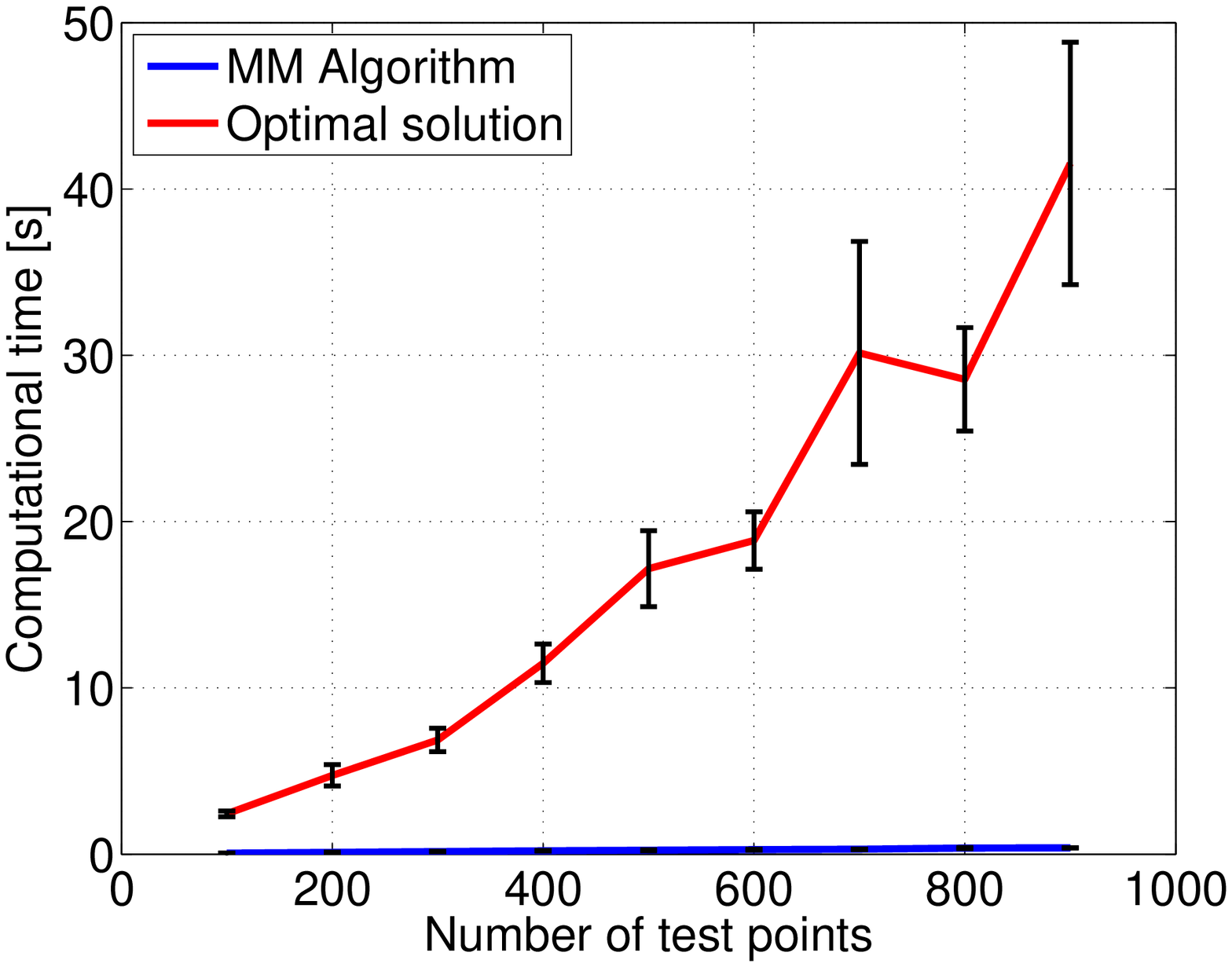}
 (b) 
 \caption{Comparison between the optimal solution of the discrete problem with the MM heuristic in a system with 100 base stations. We obtain the error bars corresponding to 95$\%$ confidence intervals by estimating the error in the mean from 100 realizations of the simulations. (a) Average number of active base stations as a function of the number test points/users. (b) Average computational time as function of the number of test points/users.} 
 \label{fig:comparison}
\end{figure*}

\section*{\small SUMMARY AND OUTLOOK}
We have showed that many communication schemes aiming at reducing the transmit energy per bit may, in fact, increase the total energy consumption. The analysis and development of wireless communication systems have traditionally considered only the energy radiated by antennas, but neglected the energy required for operating the network. By means of information-theoretic arguments, we showed that the latter cannot be neglected when considering dense deployments of base stations. One way of saving on operating power is to adjust the network to the demand by switching off unnecessary network elements. We devised algorithms to select repeatedly a suitable subset of the network's base station.

These tools for energy saving need to be refined and extended to consider, for example, mobility, temporal traffic profiles, and the high level of cooperation among network elements of future systems. In particular, many issues remain open concerning the presented energy saving optimization approach based on majorization-minimization: i) Under which conditions does the sequence produced by the majorization-minimization algorithm converge in this particular application domain? To the best of the authors' knowledge, MM-based algorithms have not been formally shown to converge even when applied to problems in compressive sensing. ii) How can we use realistic load estimates in the optimization problems (not average or worst-case estimates), since the spectral efficiency is a function of the assignments $\{x_{i,j}\}_{i\in\setm,j\in\setn}$? Integrating the results on interference calculus into the energy saving problems may be a direction for future research. iii) How are temporal traffic patterns and mobility exploited to save energy? This topic is particularly important for content-centric networking, where we have the additional option of caching content with the intent to save energy by considering current and future channel conditions. iv) How are distributed versions of the optimization algorithms implemented?  Information exchange among network elements may consume wireless resources, but this fact is not considered in most optimization models. v) How is the flexibility of choosing different communication strategies (e.g., cooperative transmission schemes) added while keeping the energy optimization problems tractable? Adding this level of flexibility complicates the optimization problems. As illustrated above, even the simpler task of computing load in the presence of cooperative systems is already difficult.

In summary, our work  shows that dense and ultra dense deployments of future 5G networks will hardly become reality unless the scaling of base stations can be decoupled from the growth of operating power. Switching off network elements when they are not required is one way to do so.
\par

\section*{\small REFERENCES}

{
\footnotesize
\bibliographystyle{IEEEtran}
\bibliography{IEEEabrv,refFP,references}
}
\end{document}